\documentclass[
reprint, 
superscriptaddress, nofootinbib, amsmath,amssymb, aps, prx,
floatfix,
]{revtex4-2}

\usepackage{siunitx}
\usepackage[normalem]{ulem}
\usepackage[inline]{enumitem}
\usepackage{mathtools}
\usepackage{graphicx}
\usepackage{dcolumn}
\usepackage{bm}
\usepackage[dvipsnames]{xcolor}
\usepackage[linkcolor=Blue,citecolor=Blue,urlcolor=Blue,colorlinks=true]{hyperref}

\DeclareMathOperator{\tr}{tr}
\usepackage{amsmath,amssymb}

\def\vec{\boldsymbol}
\def\mat{\boldsymbol}

\DeclareMathOperator{\erf}{erf}
\DeclareMathOperator{\erfc}{erfc}

\begin{document}

\title{Dynamical mean-field theory of active, self-interacting heteropolymers}%

\author{Andriy Goychuk}
\email{andriy@goychuk.me}
\affiliation{Department of Systems Immunology, Helmholtz Centre for Infection Research, 38124 Braunschweig, Germany}
\affiliation{Institute for Biochemistry, Biotechnology and Bioinformatics, Technische Universität Braunschweig, Braunschweig, Germany}
\affiliation{Lower Saxony Center for Artificial Intelligence and Causal Methods in Medicine (CAIMed), Hannover, Germany}

\date{\today}

\begin{abstract}
Analytical theories of polymer dynamics typically focus on the linear response around some steady state obtained via perturbative calculations.
The present study derives a dynamical, nonlinear mean-field theory that yields unified access to the evolution of conformations, contact probabilities, and fluctuations of active polymers within a single analytical framework.
As in previous mean-field descriptions of passive polymers, the object of interest is the two-point correlator, which is the lowest-order relevant observable in the general Langevin equation of a polymer.
In contrast, the present focus is on active polymers driven by heterogeneous monomer-level kicks and spatially correlated active flows, while allowing nonreciprocal couplings, in the presence of hydrodynamic and contact interactions.
Using a Gaussian closure to truncate higher-order correlations leads to a self-consistent diffusion equation for the chain correlations.
In some cases, this can be further simplified to the dynamics of pairwise squared separations in closed form.
In the continuum limit, the theory reveals useful analogies to similar mean-field approximations for reaction-diffusion systems, as well as fractional diffusion in different scale-free regimes.
Deriving asymptotic solutions in different parameter regimes of the continuum limit recovers the scalings of coiled, globular, and self-avoiding polymers, thus validating the theory.
Applied to hydrodynamically coupled active polymers, such as chromatin, the framework predicts a sequence of crossover scalings between coiled, swollen, and hyper-compacted fractal states.
\end{abstract}

\maketitle

\section{Introduction}%
The widespread role of polymers in industrial applications, materials science, and cell biology has sparked pioneering theoretical research ranging from scaling concepts~\cite{Book::deGennes1979} to models of polymer solutions~\cite{Book::Doi2007} and melts~\cite{Book::Fredrickson2005}, to explain spatial organization and material properties.
The physical principles linking heteropolymer sequence to organization have been investigated in energy landscape models of protein folding~\cite{Review::Onuchic1997, Review::Pande2000, Review::Shakhnovich2006, Review::Dill2008, Review::Wolynes2015, Review::Nassar2021}.
More recently, these concepts helped rationalize the spatiotemporal organization of the genome into conformational ensembles~\cite{Article::Zhang2015, Perspective::Zhang2017, Article::DiPierro2016, Article::DiPierro_2017, Article::DiPierro2018}.
This genome folding controls function in the form of gene expression by tuning the frequency of contacts among different regulatory elements~\cite{Dekker2016, Review::Misteli2020}.

In addition to passive forces, energy-consuming processes in the living cell and the cell nucleus also generate active forces that break detailed balance.
Seminal experiments have demonstrated that chromatin, consisting of genomic DNA and associated proteins, undergoes active fluctuations~\cite{Article::Weber2012} and spatiotemporally correlated flows~\cite{Article::Zidovska2013, Article::Shaban2018} with implications for function~\cite{Article::Jord2022}.
Moreover, pairwise correlated motion has been measured in locus tracking experiments~\cite{Article::Backlund2014, Article::Bruckner2023} and recently attributed to spatially correlated flows~\cite{Article::Harju_2026}.
These experimental data make chromatin a paradigmatic example of an active polymer and necessitate theoretical and analytical frameworks beyond thermal equilibrium.

Linear response theory is a powerful framework for elucidating the statistics of single-chain conformations in active and passive Rouse and wormlike chains~\cite{Article::Rouse1953, Ghosh_Gov_2014, Article::Osmanovic2017, Article::Osmanovic2018, Article::Put2019, Review::Winkler2020, Article::Goychuk2023, Article::Goychuk_2024} and Gaussian network models~\cite{Article::Sauerwald_2017, Article::LeTreut2018, Article::Shi2019, Article::Shi_Thirumalai_2021, Article::Shi_Thirumalai_2023, Article::Shi_Shin_Thirumalai_2025}.
Motivated by the coherent chromatin flows observed in experiment~\cite{Article::Zidovska2013, Article::Shaban2018}, effective linear theory enabled us to analytically relate spatiotemporally correlated motion with heteropolymer folding~\cite{Article::Goychuk2023, Article::Goychuk_2024}.
However, exploration beyond linear theory typically poses challenges due to the many coupled degrees of freedom and requires Brownian dynamics simulations~\cite{Article::DiPierro2016, Article::DiPierro_2017, Article::Brahmachari2023}, perturbative calculations, renormalization group theory, and scaling arguments~\cite{arxiv::Sakaue2026a, arxiv::Sakaue2026b}.

These challenges are not unique to polymers but are shared among classical and quantum many-particle systems.
For sufficiently large particle densities, however, interactions become self-averaging and can be approximated by mean-field theories such as the Hartree-Fock method for wave functions.
In multi-chain systems such as polymer melts, local polymer concentration profiles have been derived with self-consistent field theory~\cite{Book::Fredrickson2005}, and dynamical mean-field theory~\cite{Article::Fredrickson_Orland_2014} based on the Martin-Siggia-Rose formalism~\cite{Article::Martin_Siggia_Rose_1973}.
Analogous mean-field theories have led to major advances in understanding nonequilibrium pattern formation, such as in reaction-diffusion systems~\cite{Review::Halatek2018, Frey_Weyer_2026}.
The goal of this manuscript is to expose active polymer dynamics in sequence space to these powerful continuum limit frameworks.

In contrast to classical reaction-diffusion models, the main object of interest is not the primitive density but the correlations, or dynamical structure factor, of the polymer.
In thermal equilibrium, the corresponding partition function can be studied with the random phase approximation, which works best for polymer melts~\cite{Leibler_1980}.
Similar Gaussian approximations, by effectively modeling dynamics in mean-potential, have been used to develop dynamical mean-field descriptions of passive heteropolymers~\cite{Timoshenko_1995, Timoshenko_1995b, Timoshenko_1996, Timoshenko_1998, Jost_2014}.
In the following, motivated by these past advances and by the dense, active chromatin in the cell nucleus, I will discuss a dynamical mean-field theory for active polymers, in which fluctuations and dissipation are decoupled and pairwise interactions may be nonreciprocal.
This framework maps polymer and chromatin folding to a (fractional) diffusion problem and thereby paves the way towards analytically tractable and asymptotically solvable theories.

\section{Langevin dynamics of an active, discrete heteropolymer}%
Consider a polymer consisting of $N$ monomers, each labeled with a discrete index $i$ indicating the sequence space coordinate.
The relative arrangement of the monomers with real space coordinates $\vec{r}_i(t)$ defines the polymer conformation at time $t$.
Each chain conformation has an internal energy $E(\{\vec{r}_i(t)\})$ due to bonds between neighbors in sequence space and higher-order interactions in real space.
In addition to potential forces from the internal energy, nonequilibrium mechanisms such as reaction-diffusion-interaction can give rise to deterministic nonreciprocal forces $\vec{\chi}_i(t) \equiv \vec{\chi}_i(\{\vec{r}_i(t)\}, t)$~\cite{Article::Goh_2025}.
These nonreciprocal interactions can be derived, for example, by connecting the polymer conformation to the concentration profiles of reactants using (time-dependent) Green's functions.
Taken together, 
\begin{equation}
    \vec{f}_i(t) = - \frac{\delta E}{\delta \vec{r}_i(t)} + \vec{\chi}_i(t) \, 
\end{equation}
is the deterministic force acting on each monomer.
In addition, the monomers are agitated by stochastic forces $\vec{\eta}_i(t)$ with zero mean and covariance~\cite{Article::Goychuk2023}
\begin{equation}
\label{eq:noise_covariance_monomers}
    \langle\vec{\eta}_i(t) {\otimes} \vec{\eta}_j(t')\rangle = \frac{1}{3} \mat{I} \, C_{ij} \, \delta(t-t') \, ,
\end{equation}
where $C_{ij}$ is a symmetric and positive semi-definite matrix with units of a diffusion coefficient.
The excitations are assumed Markovian in the present work, but can contain both thermal and active components and thus break the fluctuation-dissipation theorem.
For example, diagonal entries with different magnitudes can be interpreted as loci having different levels of activity.
Off-diagonal entries could correspond to different loci showing correlated or anti-correlated responses to long-ranged stochastic forcing by electrical fields or concentration gradients~\cite{Article::Goychuk2023}.

The monomers are embedded in a fluid which mediates drag forces and hydrodynamic couplings, with mobility tensor $\mat{M}_{ij}(\{\vec{r}_i(t)\})$.
Motivated by experimental data on spatiotemporally correlated active flows in chromatin~\cite{Article::Zidovska2013, Article::Shaban2018}, the polymer is also advected with stochastic velocity $\vec{\xi}(\vec{r},t)$ with zero mean and covariance~\cite{Article::Mahajan2022, Weady_2024, Valei_2025, Article::Harju_2026}
\begin{equation}
\label{eq:noise_covariance_spatial}
    \langle\vec{\xi}(\vec{r},t) {\otimes} \vec{\xi}(\vec{r}',t')\rangle = \frac{1}{3} \mat{I} \, A \, C_\xi(\vec{r}-\vec{r}') \, \delta(t-t') \, ,
\end{equation}
on timescales larger than the experimentally measured correlation time of \SI{10}{\second}~\cite{Article::Zidovska2013}.
As will be discussed below, spatially correlated flows are, when averaged over chromatin conformations, effectively equivalent to monomer-specific correlated agitations~\cite{Article::Goychuk2023}.
Despite this mapping between sequence space and embedding space, which allows constructing analogies, both terms can generally be expected to have different physical interpretations.
To simplify the analysis, I consider monomer-specific excitations and spatial excitations to originate from different physical processes and thus as statistically independent $\langle\vec{\eta} \otimes \vec{\xi} \rangle = 0$.

Taken together, and neglecting memory in the present work by assuming that conformational changes are the slowest relevant set of degrees of freedom, the polymer obeys the following Langevin dynamics~\cite{Book::Oettinger_1996},
\begin{equation}
\label{eq:force_balance}
    \partial_t\vec{r}_i(t) = \mat{M}_{ij}(\{\vec{r}_i(t)\}) {\cdot} \Bigl[
    \vec{f}_j(t) + \vec{\eta}_j(t) \Bigr] + \vec{\xi}(\vec{r}_i(t),t) \, ,
\end{equation}
where repeated indices indicate summation.
In the following, I will systematically study Eq.~\eqref{eq:force_balance} by gradually increasing the complexity of the model for polymers in $d=3$ dimensions.

\section{Dynamical mean-field theory of polymer folding and fluctuations}

\subsection{Gaussian closure}%
A polymer such as chromatin lacking a well-defined crystal structure, up to translation and rotation, has a vanishing first moment $\langle\vec{r}_i(t)\rangle = 0$.
Such polymers only have structure in the statistical sense, represented by correlators of second order and higher even orders, while odd correlators vanish due to symmetry.
According to Marcinkiewicz' theorem~\cite{Article::Marcinkiewicz_1939}, the hierarchy of correlators either truncates at the Gaussian level, namely the two-point correlator 
\begin{equation}
\label{eq:correlation_difftime_definition}
    X_{ij}(t,t') \coloneqq \langle\vec{r}_i(t) \cdot \vec{r}_j(t')\rangle \, ,
\end{equation}
or is infinite.
Current experimental measurements on chromatin yield pairwise contact probability maps, but lack the information required to constrain higher-order correlators.
Given this lack of data, and motivated by past self-consistent mean-field modeling of passive polymers~\cite{Timoshenko_1995, Timoshenko_1995b, Timoshenko_1996, Timoshenko_1998, Jost_2014}, consider a Gaussian approximation for the joint distribution of the chain configurations.
The Gaussian distribution can be interpreted as the least-biased distribution that maximizes the information-theoretic entropy given available data, and has been used successfully to reproduce chromatin conformations in linear network models~\cite{Article::Sauerwald_2017, Article::LeTreut2018, Article::Shi2019, Article::Shi_Thirumalai_2021, Article::Shi_Thirumalai_2023, Article::Shi_Shin_Thirumalai_2025}.

The Gaussian closure is a well-known approximation to deal with the infinite Bogoliubov–Born–Green–Kirkwood–Yvon hierarchy in many-particle systems.
Truncating the series of correlators enables the use of powerful tools such as correlation splitting techniques.
By exploiting these tools, below I derive a dynamical mean-field theory for the dynamics of the position correlator 
\begin{equation}
\label{eq:correlation_sametime_definition}
    X_{ij}(t) \coloneqq \langle\vec{r}_i(t) \cdot \vec{r}_j(t)\rangle \, ,
\end{equation}
and the corresponding mean squared separation between monomers 
\begin{equation}
\label{eq:correlation_to_separation}
    \Delta R^2_{ij} \coloneqq \langle |\vec{r}_i(t) - \vec{r}_j(t)|^2 \rangle = X_{ii} + X_{jj} - 2 X_{ij} \, ,
\end{equation}
in active polymers with broken detailed balance and nonpotential forces.
For variables that depend only on a single time point, the short-hand notation $X_{ij} \equiv X_{ij}(t)$ and $\Delta R^2_{ij} \equiv \Delta R^2_{ij}(t)$ is used.

To model purely local interactions such as self-avoidance, self-attraction, or proximity ligation in the chromosome conformation capture protocol~\cite{Article::LeTreut2018}, consider the regularized contact kernel 
\begin{equation}
\label{eq:contact_probability}
    \bar{\delta}(r, a) = \left(\frac{2\pi}{3}a^2\right)^{-\frac{3}{2}} \, \exp\biggl[-\frac{3r^2}{2a^2}\biggr] \, ,
\end{equation}
with $r$ the pairwise distance and the standard deviation $a$ reflecting the typical contact range.
In the limit $a \to 0$, this converges to the 3D $\delta$-distribution.
Within the Gaussian closure, this leads to the conformation-averaged interaction probability (effective contact kernel)
\begin{equation}
\label{eq:effective_contact_probability}
    P(r,a) \equiv \left\langle \bar{\delta} \right\rangle (r, a) = \left[\frac{2\pi}{3}(a^2 + r^2) \right]^{-\frac{3}{2}} \, .
\end{equation}
In the limit $a \to 0$, this converges to the probability density $P_{ij} \equiv P(\Delta R_{ij}, 0)$ of pairwise contacts or looping in active heteropolymers such as chromatin.
Since Gaussian approximations become increasingly accurate at high concentrations in many-particle systems, I expect the mean-field theory to be applicable in the dense cell nucleus.
It will fail, however, to account for dynamics driven by rare events.

\subsection{Correlation splitting}
To simplify the theory, consider timescales much larger than the correlation time of the excitations and the memory of the solution surrounding the polymer.
This Markovian assumption can in principle be lifted because it is separate from the Gaussian approximation required for correlation splitting.
Using the Novikov-Furutsu theorem~\cite{Article::Novikov_1965, Article::Furutsu_1963}, with the excitations and the conformations (approximately) Gaussian random variables, and respecting causality of the excitations leads to (details in End Matter) the following dynamics of the same-time correlator [Eq.~\eqref{eq:correlation_sametime_definition}]:
\begin{subequations}
\label{eq:framework}
\begin{equation}
\label{eq:dynamic_mean_field_full}
    \partial_t X_{ij} 
    = \left[\hat{J} X 
    + X \hat{J}^T
    + \hat{M} C \hat{M}\right]_{ij} 
    + A \bigl\langle C_\xi \bigr\rangle (\Delta R_{ij})
    ,
\end{equation}
with $\hat{M}_{ij}$ the effective hydrodynamic kernel and $\hat{J}_{ij}$ the response matrix as detailed below.
In deriving Eq.~\eqref{eq:dynamic_mean_field_full}, I assumed fluid incompressibility and that hydrodynamic interactions are reciprocal, $\hat{M}_{ij}=\hat{M}_{ji}$, and instantaneous.
Moreover, in the last step I approximated the covariance of the mobility tensor via pre-averaging $\mat{M}_{ij} \approx \mat{I} \, \hat{M}_{ij}$.
This choice was made purely for convenience and is not strictly required to carry out the Gaussian integrals of the closure; furthermore, it does not change the asymptotic scaling (e.g., for $a \to 0$) of the effective interactions.
The last term in Eq.~\eqref{eq:dynamic_mean_field_full} reveals that the polymer fluctuations smear out the spatiotemporally correlated fluid flows, essentially amounting to a random walk in random flow~\cite{arxiv::Goychuk_2025}.

Importantly, detailed balance holds if and only if (i) interactions are reciprocal, $\hat{M}^{-1}\hat{J} \coloneqq \hat{H} = \hat{H}^T$, and (ii) the fluctuation-dissipation theorem holds, $[\hat{M} C]_{ij} = T \delta_{ij}$ with constant $T$, and (iii) $A = 0$ or $\langle C_\xi \rangle (\Delta R_{ij}) \propto \hat{M}_{ij}$.
In this passive scenario, Eq.~\eqref{eq:dynamic_mean_field_full} reduces to the Gaussian self-consistent theory derived in prior work by assuming the existence of a mean potential~\cite{Timoshenko_1995, Timoshenko_1995b, Timoshenko_1996, Timoshenko_1998, Jost_2014}.
In thermal equilibrium, the conformations of the chain $X = -\frac{1}{2}T \, \hat{H}^{-1}$ are determined purely by the energetics and the effective hydrodynamic kernel cancels out.
More broadly, this discussion reveals three distinct routes to drive the polymer out of equilibrium.
If detailed balance is broken, the conformational ensemble is not just set by the energetics, and active excitations will fold the polymer~\cite{Article::Goychuk2023}.
In summary, the dynamical mean-field theory models evolving contact maps in active polymers such as chromatin, and does not rely on the existence of a free energy.

Moreover, the theory also gives direct access to the fluctuations [Eq.~\eqref{eq:correlation_difftime_definition}], which obey 
\begin{equation}
\label{eq:dynamic_mean_field_fluctuations}
    \partial_t X_{ij}(t,t') = \left[\hat{J} X(t,t')\right]_{ij}\, , \quad \text{for} \quad t > t'\, ,
\end{equation}
\end{subequations}
with the initial condition $X_{ij}(t',t') = X_{ij}(t')$.
Note that the response matrix implicitly depends on time, $\hat{J} \equiv \hat{J}(t)$.
This expression for the fluctuations illuminates how tension propagates along the polymer and how this gives rise to the subdiffusive motion of individual loci and pairs of monomers.
In agreement with the previous discussion, it also reveals that time reversal symmetry is broken, thus indicating nonequilibrium dynamics, whenever $\hat{J} X \neq X \hat{J}^T$~\cite{submitted::Kannan_2026}.

Describing pairwise hydrodynamic interactions by the Rotne-Prager-Yamakawa tensor~\cite{Article::Rotne_Prager_1969, Article::Yamakawa_1970}, while neglecting many-body effects and using fluid incompressibility, leads to the Zimm kernel~\cite{Article::Zimm1956, Article::MartinGomez2019},
\begin{equation}
\label{eq:zimm_kernel}
    \hat{M}_{ij} \coloneqq \frac{1}{3} \left\langle \tr(\mat{M}_{ij}) \right\rangle =  
    D_0 \, h\biggl(\frac{\Delta R_{ij}^2}{6a^2}\biggr)
    \, ,
\end{equation}
with monomer radius $a$ and diffusion coefficient $D_0$.
The conformation-averaged hydrodynamic coupling, $h(r) \coloneqq \erf(r^{-1/2}) - \sqrt{r/\pi} \, (1-e^{-1/r})$, scales as $h(r) \sim (\pi r)^{-1/2}$ at large distances.
Hydrodynamic coupling causes the displacements of the chain (kicks) to be effectively correlated in sequence space even when the monomer excitations (forces) are not.
The effective response matrix is given by
\begin{equation}
\label{eq:effective_response_matrix}
    \hat{J}_{im} \coloneqq
    \hat{M}_{ij} \, \frac{1}{3} \left[ - \tr \left\langle 
    \frac{\delta^2 E}{\delta \vec{r}_m \delta \vec{r}_j}  
    \right\rangle + \left\langle \frac{\delta}{\delta \vec{r}_m} \cdot \vec{\chi}_j \right\rangle \right] \, .
\end{equation}
This reveals that the Gaussian ensemble dynamics resembles how different segments of the polymer move in response to the mean-field generated by other segments.
Consequently, the Gaussian closure is equivalent to assuming that the dynamics of each segment is fast so that it can explore the full phase space admitted by the constraints from the neighboring segments.

\subsection{Effective models based on linear response}%
In steady state, $t' \to \infty$, the conformation-averaged response matrix $\hat{J}(t')$ becomes time-invariant.
Integration of Eq.~\eqref{eq:dynamic_mean_field_fluctuations} then yields the linear response 
\begin{equation}
    X_{ij}(t,t') = \left[e^{\hat{J}(t-t')} X(t')\right]_{ij} \, .
\end{equation}
In thermal equilibrium, one has $\hat{J} = - \frac{1}{2} T \, \hat{M} X^{-1}$ which implies that different energy landscapes for polymer folding cannot be distinguished at the level of the linear response.
Characteristic signatures emerge only after applying a quench, at the level of nonlinear responses, or if the energy landscapes admit multiple stable states.
This degeneracy is actually a feature that allows constructing effective linear models.
For example, for a globule in thermal equilibrium, it implies that the effective response mimics long-ranged couplings by Hookean springs.
Such power-law couplings have been proposed by phenomenological linear models to mimic the structure and dynamical fluctuations of fractal chains~\cite{Article::Amitai_Holcman_2013, Article::Saito_Sakaue_2015, Article::Polovnikov_2018a, Article::Polovnikov_2018b}.
The Gaussian closure reveals that this is an internally consistent approach in the vicinity of the steady state, which can be determined by fixed-point iteration of Eq.~\eqref{eq:dynamic_mean_field_full}, where the linear response reproduces the relevant dynamical scaling exponents.
The framework also enables analyzing the fluctuations in general polymers beyond the steady state, and complements recent scaling arguments~\cite{GrosseHolz_2025, Article::Polovnikov_Kardar_2025, arxiv::Sakaue2026a, arxiv::Sakaue2026b}.

\subsection{Elimination of translation and swelling}
The theory can be further simplified if one constrains the response matrix in the following way.
First, assume that $\sum_i \hat{J}_{in} = \rho_1 + \mathcal{O}(1/N)$ with $\rho_1$ constant, which implies that the polymer experiences either no body force or a harmonic attraction towards the center of a confinement; the $\mathcal{O}(1/N)$ contribution could stem from nonreciprocal interactions.
Second, assume that $\sum_n \hat{J}_{in} = \rho_2$ with $\rho_2$ constant, which implies that the polymer in a hypothetical collapsed state experiences either no directed net force or a harmonic attraction toward the center of a confinement.
Together, this implies that $\rho_2 = \frac{1}{N}\sum_{in} \hat{J}_{in} = \rho_1 + \mathcal{O}(1/N) \equiv \rho$. 

The diffusivity of the center of mass of the polymer, $D \coloneqq \partial_t \langle |\vec{r}(t) - \vec{r}(t')|^2 \rangle |_{t=t'}$ with $\bar{\vec{r}} \coloneqq \frac{1}{N} \sum_i \vec{r}_i$, can be derived by using Eq.~\eqref{eq:dynamic_mean_field_full}:
\begin{equation}
\label{eq:center_of_mass_diffusivity}
    D
    = 
    \frac{1}{N^2} \sum_{ij} \left[ 
    \hat{M} C \hat{M}\right]_{ij} 
    + \frac{1}{N^2} \sum_{ij} A \bigl\langle C_\xi \bigr\rangle(\Delta R_{ij}) \, .
\end{equation}
The first term decays with the length $N$ of the polymer if the matrix of excitations $C$ is sufficiently sparse.
The summand in the second term is negligible for $\Delta R_{ij} \gg \lambda$ where $\lambda$ is the typical correlation length of the flow.
In other words, the second term also decays with $N$ if the radius of gyration of the polymer exceeds the correlation length $\lambda$ of the flows, $R_g \gg \lambda$.
In this case, the rate of diffusion of sufficiently long polymers, $N \to \infty$, will asymptotically vanish, $D\to 0$.
In contrast, if $\lambda \gg R_g$ then global advection leads to $D \to A \bigl\langle C_\xi \bigr\rangle (0)$ with a correction proportional to $R_g^2 \vec{\nabla}^2 C_\xi(0)$, which can be seen by rewriting the Gaussian mean via an exponential Laplacian [Eq.~\eqref{eq:appendix:exponential_laplacian}].

The dynamics of the radius of gyration of the polymer, $R_g^2 \coloneqq \frac{1}{2N^2} \sum_{ij} \Delta R^2_{ij}$, can be derived by using Eqs.~\eqref{eq:correlation_to_separation}~and~\eqref{eq:dynamic_mean_field_full}, leading to
\begin{equation}
\label{eq:radius_of_gyration_dynamics}
    \left(\partial_t - 2 \rho\right) R_g^2 = \frac{1}{N} \sum_i \Omega_i \, ,
\end{equation}
with
\begin{equation}
\label{eq:dynamic_mean_field_boundary}
    \Omega_i \coloneqq \left\{\hat{M} C \hat{M} + \left[A \, \bigl\langle C_\xi \bigr\rangle(0) - D\right] I - \hat{J}\Delta R^2 \right\}_{ii} \, ,
\end{equation}
where $I$ is the identity matrix.
The right-hand side of Eq.~\eqref{eq:radius_of_gyration_dynamics} is, in the continuum limit, a boundary integral over the polymer contour. 
It resembles a variational form that must be minimized with $\Omega_i$ representing a Neumann boundary condition for the mean-squared separation $\Delta R^2$.
If the polymer is not confined, $\rho \to 0$, and the radius of gyration is the slowest degree of freedom next to the center of mass, $\partial_t R_g \approx 0$, the right-hand side of Eq.~\eqref{eq:radius_of_gyration_dynamics} must vanish and $\Omega_i = 0$.
Conversely, if the polymer is sufficiently strongly confined, one can assume that folding first proceeds by bringing the radius of gyration to a steady state, $\partial_t R_g \to 0$, followed by reshuffling of the monomers to generate local structures.
This implies that $\Omega_i = \bar{\Omega}$ with constant $\bar{\Omega} = -2\rho R_g^2$.

Finally, substituting Eq.~\eqref{eq:dynamic_mean_field_full} in the time derivative of Eq.~\eqref{eq:correlation_to_separation} leads to a closed equation for the dynamics of the mean-squared separation:
\begin{multline}
\label{eq:dynamic_mean_field_separation}
    \partial_t \Delta R^2_{ij} = \left[\hat{J}\Delta R^2 + \Delta R^2 \hat{J}^T -2 \hat{M} C \hat{M}\right]_{ij} \\
    +2D - 2 A \bigl\langle C_\xi \bigr\rangle(\Delta R_{ij}) + \Omega_i + \Omega_j \, .
\end{multline}
The diffusivity $D$ of the center of mass of the polymer on short timescales cancels identically if one substitutes the boundary terms $\Omega_i$ [Eq.~\eqref{eq:dynamic_mean_field_boundary}].
The remainder of Eq.~\eqref{eq:dynamic_mean_field_separation}, with $\Omega_i = 0$ for unconfined polymers, corresponds to the bulk dynamics of the mean-squared separation.

If advection is global, with a correlation length exceeding the radius of gyration $\lambda \gg R_g$, the explicit bookkeeping of $D$ in Eq.~\eqref{eq:dynamic_mean_field_separation} will prevent diffusion of the center of mass from unphysically creating a global sink for the mean-squared separation.
Conversely, if advection is local, with a correlation length much smaller than the radius of gyration, $\lambda \ll R_g$, then $D \to 0$ prevents a global source from destabilizing the system by exciting all pairwise separations $\Delta R > \lambda$.
Thus, the separation of Eq.~\eqref{eq:dynamic_mean_field_separation} into bulk and boundary terms is important for consistently taking the continuum limit.

\subsection{Polymer response: Potential and nonpotential forces}%
\subsubsection{Reciprocal forces}
To model the energetics of the polymer, consider the following two main contributions.
First, reciprocal harmonic springs connect different monomers according to the Kirchhoff matrix $K_{ij}$ with $\sum_i K_{ij} = 0$~\cite{Article::Sauerwald_2017, Article::LeTreut2018, Article::Polovnikov_2018a, Article::Shi2019}.
Assuming only coupling among neighboring monomers, one has
\begin{equation}
    K_{ij} = k \left[\delta_{i+1,j} + \delta_{i-1,j} - \sum_n (\delta_{i+1,n} + \delta_{i-1,n}) \delta_{ij}\right] .
\end{equation}
Second, purely local interactions are incorporated via the contact kernel $\bar{\delta}(r, a)$ with $a$ the contact range [Eq.~\eqref{eq:contact_probability}].
Together, the conformational energy is given by 
\begin{equation}
    E = \sum_{i<j} \left[ \frac{1}{2} K_{ij} (\vec{r}_i - \vec{r}_j)^2 + \gamma \, \bar{\delta}(|\vec{r}_i - \vec{r}_j|, a)\right] \, ,
\end{equation}
up to the second virial coefficient $\gamma$~\cite{Book::Doi2007}.
Here, $\gamma > 0$ corresponds to repulsive interactions indicative of self-avoidance, whereas $\gamma < 0$ models pairwise attraction upon contact.
After evaluating the Gaussian integrals, one has
\begin{subequations}
\label{eq:effective_response_matrix_eval}
\begin{equation}
\label{eq:effective_response_matrix_compact}
    - \frac{1}{3} \tr \left\langle 
    \frac{\delta^2 E}{\delta \vec{r}_i \delta \vec{r}_j}  
    \right\rangle = 
    K_{ij} + S_{ij} \, ,
\end{equation}
with the self-interactions
\begin{equation}
\label{eq:effective_response_matrix_self}
    S_{ij} = - 2\pi \left[\gamma P^{\frac{5}{3}} (\Delta R_{ij}, a)  
    - \sum_n \gamma P^{\frac{5}{3}}(\Delta R_{nj}, a) \, \delta_{ij} \right] ,
\end{equation}
\end{subequations}
forming a Kirchhoff matrix.
These local interactions can be made specific to pairs of monomers by promoting $\gamma$ to a symmetric matrix.

The second term in Eq.~\eqref{eq:effective_response_matrix_self} is an effective local pressure which ensures that self-avoidance is reciprocal and which pushes monomers away from the centroid of the polymer for $\gamma>0$.
Compared to the contact probability $P$, Eq.~\eqref{eq:effective_contact_probability}, the response matrix due to contacts shows an additional scaling $\propto P^{2/3}$.
This is because the Hessian [cf. Eq.~\eqref{eq:effective_response_matrix_compact}] involves real-space gradients averaged over the chain conformation, and these distances scale as $P^{-1/3}$.
This interpretation is consistent with the idea that each polymer segment responds to the mean-field generated by other segments.
It also allows extrapolating that $n$-body interactions would enter as $P^{2/3} \color{gray}\times\color{black} P^{n-1}$ with an additional sum over particle combinations.

\subsubsection{Nonreciprocal forces from chemical reactions}
Finally, the model can incorporate nonreciprocal interactions mediated by, for example, transcribed RNA, as we have recently investigated using Brownian dynamics simulations~\cite{Article::Goh_2025}.
Assuming that RNA diffuses quickly compared to chromatin fluctuations, the 3D steady-state concentration profile due to RNA synthesis at a monomer $i$, diffusion, and degradation or processing is given by
\begin{equation}
\label{eq:nonreciprocal_source}
    c(\vec{r} - \vec{r}_i) \propto \frac{e^{-|\vec{r} - \vec{r}_i| / \ell_i}}{|\vec{r} - \vec{r}_i|} \, ,
\end{equation}
with $\ell_i$ the diffusion length of the specific RNA species.
Gradients in this concentration profile generate forces that act on other monomers, potentially mediated by biomolecular condensates, and can be approximated by~\cite{Article::Goh_2025}
\begin{equation}
\label{eq:nonreciprocal_gradient_force}
    \vec{\chi}_j = - \sum_i B_{ji} \, \frac{\delta}{\delta \vec{r}_j} \frac{e^{-|\vec{r}_j - \vec{r}_i| / \ell_i}}{|\vec{r}_j - \vec{r}_i|} \, ,
\end{equation}
where the non-symmetric matrix $B_{ji}$ indicates how strongly monomer $j$ is attracted (negative values) or repelled (positive values) by the typical amount of RNA produced at monomer $i$.
This amounts to nonreciprocal, screened Coulomb interactions.
Their contribution to the effective response matrix, Eq.~\eqref{eq:effective_response_matrix}, is given by
\begin{equation}
    \left\langle \frac{\delta}{\delta \vec{r}_i} \cdot \vec{\chi}_j \right\rangle = 
    \frac{B_{ji}}{\ell_i^3} \, b\biggl(\frac{\Delta R_{ij}^2}{6\ell_i^2}\biggr)
    -
    \sum_n \frac{B_{jn}}{\ell_n^3} \, b\biggl(\frac{\Delta R_{nj}^2}{6\ell_n^2}\biggr)  \delta_{ij} \, ,
\end{equation}
with $b(r) \coloneqq - (4\pi r^3)^{-1/2} + (\pi r)^{-1/2} - e^{r} \erfc(r^{1/2})$ scaling as $b(r) \sim -(4\pi r^3)^{-1/2}$ for $r \to 0$ which corresponds to large diffusion lengths.
Note that the function $b(r)$ is strictly negative and increasing monotonically with $r$.
The above long-range interaction terms diverge at contact, $\Delta R_{ij}^2 \to 0$ because of the point source in Eq.~\eqref{eq:nonreciprocal_source} and because the gradient is evaluated at a single point in Eq.~\eqref{eq:nonreciprocal_gradient_force}.
In simulations, this must be regularized, $\Delta R_{ij}^2 \to a^2 + \Delta R_{ij}^2$, which effectively corresponds to a cutoff at the finite monomer size $a$.

\begin{figure}[t]
    \centering
	\includegraphics{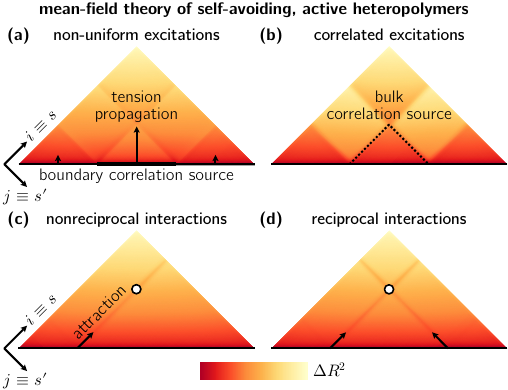}
	\caption{%
    \textbf{Effective diffusion of pairwise squared separations and correlations in the dynamical mean-field theory}.
    Pairwise squared separation $\Delta R^2$ for a discrete heteropolymer consisting of $N=1000$ monomers that repel each other (virial coefficient $\gamma = 1$), are connected by nearest-neighbor springs (stiffness $k = 1/2$), and experience thermal excitations (diffusion coefficient $T = 1$).
    The Gaussian pairwise contact probability is $P = (2\pi/3)^{-3/2} \Delta R^{-3}$.
    (a) Increasing the magnitude of excitations in the central segment (by $\times 7.5$) locally stretches the chain as reported in Ref.~\cite{Article::Goychuk2023}.
    The underlying mechanism is an increased boundary source (influx) of correlations, equivalent to a sink for the mean squared separation.
    (b) Correlated excitations ($C_{ij}$ increased by $+ 0.5$ in the dashed region) locally condense the chain as reported in Ref.~\cite{Article::Goychuk2023}.
    The underlying mechanism is a bulk source of correlations, equivalent to a bulk sink for the mean squared separation.
    (c) Molecules produced at a locus $i = 0.75 N$ non-reciprocally attract the locus $j = 0.25 N$ with interaction parameter $B_{ji} = -25$, following the condensate-based mechanism discussed in Ref.~\cite{Article::Goh_2025}.
    This leads to a characteristic line in the squared separation and contact maps.
    (d) If the long-ranged interactions are reciprocal, $B_{ji} = B_{ij} = -25$, one finds two characteristic lines.
    Pairwise squared separation maps are shown at $t = 1000$ after numerically integrating Eq.~\eqref{eq:dynamic_mean_field_separation}.
    The system was initialized at $t = 0$ with the conformation of an equilibrated Rouse chain.
    All interactions were cut off at the monomer size $a = 1$.
	}
	\label{fig::triangular_domain_diffusion}
\end{figure}

Figure~\ref{fig::triangular_domain_diffusion} shows simulations of the specific model discussed in this section.
These calculations recover our prior results based on linear response theory and Brownian dynamics simulations for (a) non-uniform and (b) correlated excitations~\cite{Article::Goychuk2023}.
The structure of the dynamical mean-field theory with sources of correlations on the boundary ($i=j$ for discrete chains, $s=s'$ in the continuum limit) and in the bulk ($i \neq j$ for discrete chains, $s \neq s'$ in the continuum limit) is independent of nonlinearities such as self-avoidance.
This explains why the linear response theory was so successful in predicting Brownian dynamics simulations~\cite{Article::Goychuk2023}.
If biomolecules such as RNA are produced at a specific locus $i \equiv s$, such as a promoter, the resulting gradients can attract a different locus $j \equiv s'$, such as an enhancer with a bound biomolecular condensate~\cite{Article::Goh_2025}.
The theory predicts that in contact maps, such nonreciprocal interactions show as a (c) contact line that is non-symmetric about the normal vector on the boundary.
In contrast, reciprocal interactions show as (d) two contact lines that are symmetric about the normal vector on the boundary and intersect at $i=j$.
Although the inverse statement does not automatically follow, because different mechanisms can create similar conformations, this prediction can be tested by digestion of RNA.

\section{Continuum limit reveals diffusion on a triangular domain}%
The discussion so far has developed a mean-field theoretical framework for studying the dynamics of active polymers, which generalizes previous self-consistent Gaussian approximations of passive polymers~\cite{Timoshenko_1995, Timoshenko_1995b, Timoshenko_1996, Timoshenko_1998, Jost_2014}, and has been centered around a discrete chain.
To extract further insights, now consider the continuum limit by mapping discrete monomer indices to a continuous sequence space, $i \mapsto s/ds$, where the discretization $ds$ is taken equivalent to the contact range or the monomer radius, $ds \equiv a$.
This mapping turns the correlation and mean squared separation matrices in Eqs.~\eqref{eq:dynamic_mean_field_full},~\eqref{eq:dynamic_mean_field_fluctuations},~and~\eqref{eq:dynamic_mean_field_separation} into continuous functions such as $X_{ij} \mapsto X(s,s')$ and $\Delta R^2_{ij} \mapsto \Delta R^2(s,s')$.
In this formalism, the response of the system is described by the operator $\hat{J}_{ij} \mapsto \hat{\mathcal{J}}(s,s')$ which characterizes tension propagation via different mechanisms.
Finally, tensor contractions over a monomer index become convolutions in sequence space, e.g., $\hat{J}_{in} X_{nj} \mapsto \int dx \hat{\mathcal{J}}(s,x) X(x,s')$, in short $\hat{\mathcal{J}} X(s,s')$, where the placement indicates what variable the operator acts on.
This operator formalism will enable a more intuitive discussion and, moreover, a mapping to fractional diffusion in different self-consistent scaling regimes, which will enable operator classification.
Conceptually, it reframes the dynamics of polymer conformations as a 2D diffusion problem that represents the propagation of tension in sequence space (Fig.~\ref{fig::triangular_domain_diffusion}).

\subsection{Rouse polymer}
To gradually build intuition, first consider the simplest possible scenario of a Rouse chain, with interaction matrix $K_{ij} = k \, [\delta_{i+1,j} + \delta_{i-1,j} - 2\delta_{ij}]$ corresponding to Hookean springs between neighboring monomers, in absence of spatial excitations, $A \to 0$, self-interactions, $\gamma \to 0$, and hydrodynamic interactions, $\hat{M}_{ij} \to \delta_{ij}$.
In the continuum limit, the springs between neighbors correspond to a Laplace operator, leading to the response operator $\hat{\mathcal{J}}(s,s') = \kappa \delta(s-s') \partial_{s'}^2$, in short $\hat{\mathcal{J}} = \kappa \partial_s^2$, with tension $\kappa = k \, ds^2$.
Thus, Eq.~\eqref{eq:dynamic_mean_field_full} can be interpreted as a diffusion equation 
\begin{equation}
\label{eq:dynamic_mean_field_correlation_rouse}
    \partial_t X(s,s') = \kappa \, [\partial_s^2 + \partial_{s'}^2] X(s,s') + C(s,s') \, .
\end{equation}
Since the continuum limit corresponds to the number of monomers $N\to\infty$, the equivalent of Eq.~\eqref{eq:dynamic_mean_field_separation} is given by
\begin{equation}
\label{eq:dynamic_mean_field_separation_rouse}
    \partial_t \Delta R^2(s,s') = \kappa \, [\partial_s^2 + \partial_{s'}^2] \Delta R^2(s,s') - 2C(s,s') \, .
\end{equation}
This reveals in a transparent way, without resorting to Fourier transforms or mapping to an equivalent equilibrium system, how correlated excitations reduce the mean squared separation between different loci~\cite{Article::Goychuk2023}.
If the springs between neighbors can vary in stiffness, then the 1D response operator will be modified to a Sturm-Liouville form, $\hat{\mathcal{J}} = \partial_s \kappa \partial_s$, incorporating space-dependent diffusion.

Due to symmetry under relabeling $s \leftrightarrow s'$, the relevant domain for this diffusion process can be considered as the triangular half-space defined by $s \geq s'$.
Spatially independent but nonuniform excitations, $C(s,s') = C(s) \delta(s-s')$ lead to a Neumann boundary condition, 
\begin{equation}
    \vec\nabla X(s,s')|_{s=s'} = - \frac{C(s')}{2\kappa} \, \frac{1}{\sqrt{2}}\hat{\vec{e}}_{s-s'} \, ,
\end{equation}
for the slope of the two-point correlation function.
This can be derived by integrating Eq.~\eqref{eq:dynamic_mean_field_correlation_rouse} along an infinitesimal segment perpendicular to the diagonal ($s=s'$) and taking into account that $X(s,s') \sim X(s-s')$ for $s \to s'$.
It also reveals that correlations are injected at the diagonal ($s=s'$) and propagated away by tension, which explains how patterns imposed by nonuniform activity emerge~\cite{Article::Goychuk2023}.
For the mean squared separation, the Neumann boundary condition reads
\begin{equation}
    \vec\nabla \Delta R^2(s,s')|_{s=s'} = \frac{C(s')}{2\kappa} \, \sqrt{2} \hat{\vec{e}}_{s-s'} \, .
\end{equation}
This mapping turns the uniform, and even the nonuniform, Rouse model into a fairly simple initial value problem.
For example, for uniform excitations $C(s) = C_0$, integration immediately leads to $\Delta R^2(s,s') = \frac{C_0}{2\kappa} \, |s-s'|$ in steady state.
For nonuniform excitations $C(s)$, one can use a separation-of-variables ansatz to recover the Green's functions discussed in Ref.~\cite{Article::Goychuk2023}.

\subsection{Zimm polymer}
Next, consider a polymer with hydrodynamic interactions according to the Zimm kernel [Eq.~\eqref{eq:zimm_kernel}] and nearest neighbor springs as before.
Taking the monomer radius $a \equiv ds \to 0$ according to the continuum limit leads to
\begin{equation}
    \hat{\mathcal{J}} X(s,s') = \sqrt{\frac{6}{\pi}} D_0 \kappa \, \int_{-\infty}^{\infty}\!\! dx \frac{\partial^2_{x} X(x,s')}{\Delta R(s,x)} \, ,
\end{equation}
for an infinitely long polymer.
Now, consider a locally homogeneous and scale-free regime of the chain in which $\Delta R(s,x) = L^{1-\nu} |s-x|^\nu$ with $L$ some local lengthscale,
\begin{equation}
    \hat{\mathcal{J}} X(s,s') = \sqrt{\frac{6}{\pi}} D_0 \kappa \frac{L^\nu}{L} \, \int_{-\infty}^{\infty}\!\! dx \frac{\partial^2_{x} X(x,s')}{|s-x|^\nu} \, .
\end{equation}
With $0 < 1 - \nu < 1$, this can be rewritten using the Riesz potential from fractional calculus~\cite{Review::Samko_2013}, which effectively defines an inverse fractional Laplacian, and leads to the following response operator, written in short form,
\begin{equation}
\label{eq:response_operator_zimm}
    \hat{\mathcal{J}} = D_0 \frac{\sqrt{6\pi}}{\Gamma(\nu) \cos\frac{\pi\nu}{2}} \, 
    \left(-L^2\partial_s^2\right)^{\frac{\nu-1}{2}} \kappa \partial^2_{s} \, ,
\end{equation}
where $\Gamma$ is the Gamma function.
Equation~\eqref{eq:response_operator_zimm} indicates a composition of the hydrodynamic mobility operator and the tension operator.
This reveals that hydrodynamic interactions can be interpreted as fractional diffusion of tension.
For the passive Zimm model whose steady-state conformations match the Rouse chain, $\nu = 1/2$, this immediately recovers tension propagation of wave modes $q$ according to $q^{3/2}$ in absence of self-interactions~\cite{Book::Doi2007}.

\subsection{Self-interacting polymer}

Finally, consider self-interactions according to Eq.~\eqref{eq:effective_response_matrix_self}.
Taking the monomer radius $a \equiv ds \to 0$ according to the continuum limit leads to [cf. effective contact kernel Eq.~\eqref{eq:effective_contact_probability}]
\begin{equation}
\label{eq:interaction_pairwise}
    \mathcal{S} X(s,s') = - \frac{3\gamma}{a} \left(\frac{3}{2\pi}\right)^{\frac{3}{2}} \int_{-\infty}^{\infty}\!\! dx \frac{X(x,s') - X(s,s')}{\Delta R^5(s,x)} ,
\end{equation}
for an infinitely long polymer.
Now, consider a locally homogeneous and scale-free regime of the chain in which $\Delta R(s,x) = L^{1-\nu} |s-x|^\nu$ with $L$ some local lengthscale,
\begin{equation}
\label{eq:interaction_pairwise_scalefree}
    \mathcal{S} X(s,s') = - \frac{3\gamma}{a} \frac{L^{5\nu}}{L^{5}} \left(\frac{3}{2\pi}\right)^{\frac{3}{2}} \int_{-\infty}^{\infty}\!\! dx \frac{X(x,s') - X(s,s')}{|s-x|^{5\nu}} .
\end{equation}
For $0 < 5\nu - 1 < 2$, this can be rewritten using the Riesz fractional Laplacian~\cite{Review::Samko_2013}, which leads to the following effective self-interaction operator, written in short form,
\begin{equation}
\label{eq:interaction_operator_pairwise_scalefree}
    \mathcal{S} = - \frac{3\gamma}{a} \frac{1}{L^{4}} \left(\frac{3}{2\pi}\right)^{\frac{3}{2}} \frac{\pi}{\Gamma(5\nu) \cos\frac{5\pi\nu}{2}} \left(-L^2\partial_s^2\right)^{\frac{5\nu-1}{2}} .
\end{equation}
Here, several interesting comments are in order.
For $2/5 \leq \nu < 3/5$, the hypersingular integral defining the fractional Laplacian is evaluated by calculating its principal value.
In the limit approaching the Flory scaling of a self-avoiding chain, $\nu \to 3/5$, it converges to the regular Laplacian, but requires additional regularization to be integrable for $\nu \geq 3/5$.
The reason is that, for $\nu \to 3/5$, Eq.~\eqref{eq:interaction_pairwise_scalefree} diverges logarithmically as $\log N$ unless there is a short-distance cutoff (monomer size $a$) and a long-distance cutoff (polymer length $N a$).
This divergence can only be absorbed by asymptotically weakening polymer self-avoidance, $\gamma/a \sim 1/\log N \to 0$.

Nonetheless, the self-interaction operator remains well-behaved even without such regularization for the exact scaling of a self-avoiding chain, $\nu = 0.588$.
If one singles out the propagation of contacts alone, and neglects tension, for example when the polymer is initialized in an overcompacted nonequilibrium state, then the theory predicts a dynamical scaling of $q^{5\nu - 1}$ with wave mode $q$ for free-draining polymers and, using operator composition with hydrodynamic interactions, $q^{6\nu - 2}$ for Zimm chains.
Importantly, this analysis is separate from the dynamical fluctuations of polymers around their steady state, which should be evaluated with Eq.~\eqref{eq:dynamic_mean_field_fluctuations} yielding different dynamical scalings, namely $q^{1+2\nu}$ for free-draining polymers and $q^{3\nu}$ for Zimm chains.

\begin{figure*}[ht]
    \centering
	\includegraphics{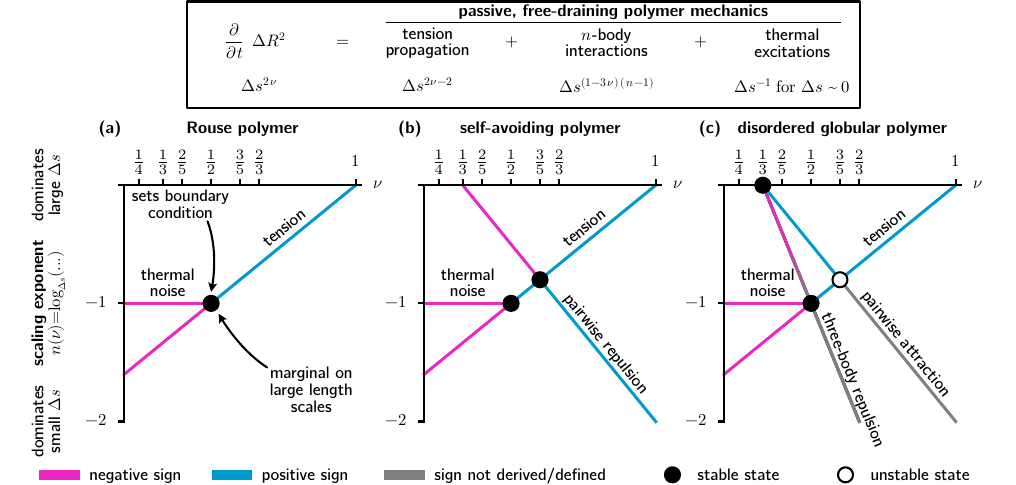}
	\caption{%
    \textbf{Asymptotic analysis of a passive polymer}.
    Hydrodynamic interactions are irrelevant for the equilibrium steady state and were therefore neglected.
    The box depicts the different competing terms in the dynamical mean-field theory [Eq.~\eqref{eq:dynamic_mean_field_separation}] and their asymptotic behaviors.
    Lines indicate the scaling exponents of the different labeled terms, assuming that the mean-squared separation has a typical exponent $2\nu$, and their sign (magenta indicates negative, blue indicates positive).
    Stable states require lines of different signs to meet.
    (a) The Rouse chain is the simplest model of a polymer.
    For $\Delta s \sim 0$, the competition between tension and thermal excitations defines $\Delta R^2 \sim \Delta s$ and sets the Kuhn length.
    For $\Delta s \gg 0$, there is no additional term that could compete with the tension, which therefore self-tunes to the marginal state $\nu = 1/2$.
    (b) For self-avoiding polymers, pairwise repulsion competes with tension and therefore leads to the classical Flory scaling $\nu = 3/5$ at the level of power counting.
    The pairwise interaction term scales as $\log N$ at the Flory fixed point, and thus diverges for infinitely long continuous polymers.
    Hence, this fixed point can only be approached from below unless the interaction parameter is asymptotically rescaled (regularized).
    (c) Self-attracting polymers collapse to a dense globular state with $\nu = 1/3$, which is stabilized by three-body interactions.
    Note that the pairwise interaction term also has a pole for $\nu \to 1/3$, which however can be balanced by higher-order interactions.
	}
	\label{fig::scaling}
\end{figure*}

\section{Asymptotic solutions}

\subsection{Flory scaling, chain collapse, and globule}%
By combining the Gaussian closure with fractional calculus, the previous section effectively linearized hydrodynamic interactions and self-interactions of polymers in approximately homogeneous, scale-free regimes.
The payoff of this analysis is that it enables an analytical derivation of the asymptotic behavior of the mean-squared separation $\Delta R^2(s,s') = L^{2-2\nu} \, |s-s'|^{2\nu}$, where $\Delta s \coloneqq |s-s'|$, and a self-consistent selection of the corresponding scaling exponent $\nu$.
To validate the theory, first consider how it recovers classical results such as the Flory scaling of self-avoiding polymers~\cite{Article::Flory_1949}, which was originally based on arguments for the primitive probability density, in thermalized polymers with contact energy $\gamma \neq 0$ and with reciprocal couplings.
In comparison to previous work recovering these scalings via spectral analysis of a mean-field theory of passive polymers~\cite{Timoshenko_1995b}, the sequence-space analysis below gives access to the sign of each term and thus, beyond power counting, allows discussing stability.

The asymptotic solution of Eq.~\eqref{eq:dynamic_mean_field_separation} requires identifying and balancing the dominant terms according to their magnitude and sign.
In this section, I will neglect hydrodynamic interactions because they do not affect the steady state of passive polymers.
The tension term is 
\begin{equation}
\label{eq:tension_scaling}
    \kappa (\partial_s^2 + \partial_{s'}^2) \Delta R^2 = 4\kappa \nu (2\nu - 1) \left|\frac{s-s'}{L}\right|^{2\nu-2} \, ,
\end{equation}
and becomes marginal for Rouse chains, $\nu = 1/2$.
Near $\Delta s \sim 0$, the kink in the mean squared separation causes the tension term to scale as $\sim +\Delta s^{2\nu-2}$ and balance the excitations $\sim -\Delta s^{-1}$, again consistent with Rouse behavior and with directly integrating Eq.~\eqref{eq:dynamic_mean_field_separation_rouse}.
Pairwise contacts contribute [Eq.~\eqref{eq:interaction_operator_pairwise_scalefree}] 
\begin{multline}
\label{eq:contact_scaling}
    \mathcal{S}\Delta R^2 + \Delta R^2 \mathcal{S} = \frac{6\gamma}{a} \frac{1}{L^{2}} \left(\frac{3}{2\pi}\right)^{\frac{3}{2}} 
    \left(\frac{\cos\frac{\pi\nu}{2}}{\cos\frac{5\pi\nu}{2}} - 1\right)
    \color{gray}\times\color{black} \\ \color{gray}\times\color{black}
    \mathrm{B}(2\nu+1, 3\nu-1) 
    \left|\frac{s-s'}{L}\right|^{1-3\nu} ,
\end{multline}
for $\nu > 1/3$, where $\mathrm{B}$ is the Beta function.
The scaling of this pairwise interaction term can also be extracted from Eq.~\eqref{eq:interaction_pairwise}, and generalized to $n$-body interactions based on the following argument.

Within the Gaussian approximation, one can factorize $n$-body interactions into $n-1$ pairwise contacts.
The probability of each pairwise contact scales as the volume explored by the chain, $P \sim \Delta R^{-3} \sim \Delta s^{-3\nu}$.
An operator that weighs the contact energies for $n$ contacts requires self-convolving $P \ast P \ast \dots \ast P$ where the contact probability appears $n-1$ times.
Together with the convolution required for the action of the operator, one has a scaling $\Delta s^{n-1}$ from the convolutions, and a scaling $P^{n-1}$ from the contacts.
Finally, taking the Hessian of the contact energy contributes $\Delta R^{-2} \sim \Delta s^{-2\nu}$, which is canceled by the scaling of the mean-squared separation $\Delta R^{2} \sim \Delta s^{2\nu}$ that the operator acts on.
Together, this means that for $n$-body interactions, the corresponding term in the dynamical mean-field theory will scale as $\gamma_n \Delta s^{(n-1)(1-3\nu)}$ with $\gamma_n$ the corresponding virial coefficient.
This classification of the operators and the corresponding terms in the dynamical mean-field theory allows analyzing which asymptotic scaling exponent $\nu$ is realized in different regimes (Fig.~\ref{fig::scaling}), as will be explained next.

For nearby monomers, $\Delta s \to 0$, the combination of terms with opposite signs and the steepest scaling dominates $\partial_t \Delta R^2(s,s')$ [cf. Eq.~\eqref{eq:dynamic_mean_field_separation}, box in Fig.~\ref{fig::scaling}] and must therefore balance to prevent divergence.
In this regime, as mentioned above, the Rouse scaling $\nu = 1/2$ is stable because tension balances excitations, as shown schematically in Fig.~\ref{fig::scaling}(a), as well as repulsive ternary interactions.
For distant monomers, $\Delta s \to \infty$, purely local excitations and terms with the steepest scaling vanish.
Instead, the combination of terms with opposite signs and the shallowest scaling must balance.
For a Rouse chain, because there is no other term to compete with, the tension self-tunes to $\nu = 1/2$ where it becomes marginal [Fig.~\ref{fig::scaling}(a)].
For self-avoiding polymers with $\gamma > 0$, regularized so that $\gamma \log N$ is finite, the classical Flory scaling $\nu = 3/5$~\cite{Book::Doi2007} is stable because tension balances repulsive binary interactions, as shown in Fig.~\ref{fig::scaling}(b).
Finally, self-attracting polymers with $\gamma < 0$ collapse to a disordered (Lifshitz) globule $\nu = 1/3$~\cite{Review::Lifshitz_Grosberg_Khokhlov_1978}, where attractive binary interactions are balanced by repulsive ternary and higher order interactions, as shown in Fig.~\ref{fig::scaling}(c).
This asymptotic analysis can be generalized from 3D to higher dimensions.

In addition to recovering these textbook results, the theory also allows one to deduce which regimes cannot be realized and why.
Because tension according to Eq.~\eqref{eq:tension_scaling} is marginal in the Rouse case, $\nu = 1/2$, it is always destabilized by pairwise repulsive or attractive interactions.
Similarly, $\nu = 2/3$ indicative of a competition between local excitations and self-attraction, is destabilized by finite tension which causes the polymer to collapse.
Finally, for self-avoiding polymers with non-regularized $\gamma > 0$, see discussion below Eq.~\eqref{eq:interaction_operator_pairwise_scalefree}, the polymer can only approach the Flory scaling from below because the contact operator becomes singular for $\nu = 3/5$.
Together, these results indicate that the dynamical theory derived from a Gaussian closure that truncates the correlator hierarchy at pairwise interactions is consistent with classical equilibrium polymer literature.
In the next section, this non-perturbative method will be used to study the dynamics of active heteropolymers in the presence of hydrodynamic interactions, which is widely considered analytically challenging.

\begin{figure*}[ht]
    \centering
	\includegraphics{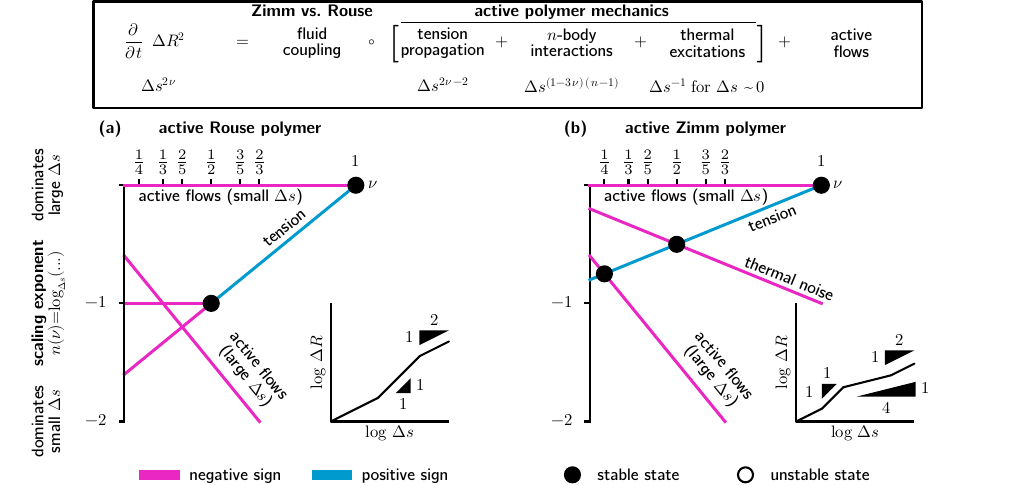}
	\caption{%
    \textbf{Asymptotic analysis of an active polymer}.
    The box depicts the different competing terms in the dynamical mean-field theory [Eq.~\eqref{eq:dynamic_mean_field_separation}] and their asymptotic behaviors.
    Lines indicate the scaling exponents of the different labeled terms, assuming that the mean-squared separation has a typical exponent $2\nu$, and their sign (magenta indicates negative, blue indicates positive).
    Stable states require lines of different signs to meet.
    Insets depict predicted behavior of the root-mean-squared separation $\Delta R(s,s')$ between monomers $s$ and $s'$.
    (a) Active fluid flows cause an effective straightening and swelling of Rouse chains on intermediate lengthscales.
    Conversely, it appears as if the regions receiving correlated excitations are more compact than the regions receiving independent kicks.
    (b) Hydrodynamic coupling changes the scaling and sign behavior of the different competing terms in the dynamical mean-field theory of Zimm polymers.
    This enables active fluid flows to create a new stable fixed point $\nu = 1/4$ that would be prohibited in absence of hydrodynamic coupling.
	}
	\label{fig::scaling_active}
\end{figure*}

\subsection{Correlated random flows can stretch or compact polymers}%
Finally, motivated by experimental measurements of coherent chromatin motion~\cite{Article::Zidovska2013, Article::Backlund2014, Article::Shaban2018}, I will study how polymer conformations can be reorganized by spatially correlated advection.
Active flows in the cell nucleus decorrelate on a timescale of seconds~\cite{Article::Zidovska2013} and thus can be considered approximately Markovian for sufficiently slow conformational changes.
For the spatial profile of the active flows, consider finite-range correlations $C_\xi(\vec{r}) = \bar{\delta}(|\vec{r}|, \lambda)$ [cf. Eq.~\eqref{eq:contact_probability}].
The micron-scale correlation length $\lambda$~\cite{Article::Zidovska2013} far exceeds the range of molecular contacts, $\lambda \gg a$.
It defines a typical crossover distance $\Delta R \sim \lambda$ between different polymer behaviors.

Due to the large length of chromatin, I assume that its center-of-mass diffusivity induced by coherent active flows is small, $D \to 0$.
The additional growth rate of the mean-squared separation according to Eq.~\eqref{eq:dynamic_mean_field_separation} is given by [cf. Eq.~\eqref{eq:effective_contact_probability}]:
\begin{equation}
    \bigl\langle C_\xi \bigr\rangle (\Delta R) = \left[\frac{2\pi}{3}\left(\lambda^2 + \Delta R^2\right) \right]^{-\frac{3}{2}} \, .
\end{equation}
As in the last sections, I will study locally homogeneous and scale-free regimes of the polymer in which $\Delta R(s,s') = L^{1-\nu} |s-s'|^\nu$ with $\Delta s \coloneqq |s-s'|$.
The conformation-averaged correlation function of the active flows asymptotically approaches
\begin{equation}
\label{eq:flow_correlations_asymptotic}
    \bigl\langle C_\xi \bigr\rangle = \left(\frac{3}{2\pi}\right)^{\frac{3}{2}} \color{gray}\times\color{black}
    \begin{dcases}
    \frac{1}{\lambda^{3}} \, , & \text{for} \, \Delta R \ll \lambda \, , \\
    \frac{1}{L^3} \left|\frac{s-s'}{L}\right|^{-3\nu} \, , & \text{for} \, \Delta R \gg \lambda \, .
    \end{dcases}
\end{equation}
In principle, one can use the logic outlined below to also consider power-law correlated flows $C_\xi(\vec{r}) \sim 1/|\vec{r}|^n$, which scale as $\Delta s^{-n \nu}$ and are thus equilibrium-like for $n=1$, finite-range-like for $n = 3$, and require regularization to prevent a UV divergence for $n \geq 3$.

\subsubsection{Actively stirred Rouse polymers effectively straighten}

For simplicity, in this section I will neglect hydrodynamic couplings and self-interactions among the monomers of the chain.
In this case, there are only two terms that can potentially compete on large lengthscales, namely tension and active fluid flows.
Naive matching of the scaling exponents suggests an asymptotic scaling $\nu = 2/5$ for finite-range correlated flows, which is more compact than a Rouse chain but less compact than a globule~\cite{Review::Lifshitz_Grosberg_Khokhlov_1978, Article::Grosberg_1993}. 
However, as shown in Fig.~\ref{fig::scaling_active}(a), this fixed point cannot be realized because the tension of a free-draining polymer changes sign for $\nu < 1/2$ so that it matches the sign of the flow correlations.
Consequently, on large lengthscales, the polymer conformations resemble a Rouse chain with $\nu = 1/2$. 
For monomers $s$ and $s'$ that are sufficiently close in sequence space, the flow correlations saturate [Eq.~\eqref{eq:flow_correlations_asymptotic}] and admit a fixed point $\nu = 1$ where the chain is effectively straightened.
Finally, for $|s-s'| \to 0$, the polymer again recovers Rouse behavior $\nu = 1/2$ due to the Neumann boundary condition.

If chromatin undergoes spatially correlated nonequilibrium excitations as recently inferred from pairwise locus tracking experiments and simulated with Rouse phantom chains lacking hydrodynamic coupling~\cite{Article::Harju_2026}, then this structural signature should be visible in contact frequency data.
Interestingly, note that excitations correlated in real space map, via the Gaussian approximation and mean squared separation, to effective excitations correlated along the sequence of the chain.
Thus, the theory predicts that the regions $|s-s'| \to 0$ experiencing correlated excitations are more compact, in the sense of a smaller slope of $\partial_s \Delta R^2$, than distant regions experiencing uncorrelated kicks.
This is consistent with linear response theory and simulations in our previous work, where correlated kicks led to an effective attraction among monomers independent of distance in sequence space~\cite{Article::Goychuk2023}.

\subsubsection{Actively stirred Zimm polymers appear to undergo hypercompaction}

The structures of a passive Zimm chain and a passive Rouse chain are identical in thermal equilibrium.
However, these two models differ in how tension propagates and in how the passive steady state is achieved, with fundamental implications for the dynamics of active polymers.
To illustrate this point, first consider thermal excitations, which are purely local in the Rouse model, $\hat{M} C \hat{M} = T I \equiv (T a) \, \delta(s-s')$ with $I$ the identity operator, but are non-local for a Zimm polymer [Eq.~\eqref{eq:zimm_kernel}],
\begin{equation}
    \hat{M} C \hat{M} = T \hat{M} \equiv (T a) \, \sqrt{\frac{6}{\pi}} D_0 \, \frac{1}{\Delta R} \, ,
\end{equation}
in the continuum limit $a\to 0$.
Assuming a locally homogeneous and scale-free regime of the polymer as before, one has
\begin{equation}
    \hat{M} C \hat{M} = \frac{T a}{L} \, \sqrt{\frac{6}{\pi}} D_0 \, \left| \frac{s-s'}{L} \right|^{-\nu} \, .
\end{equation}
This nonlocality implies that thermal excitations generally cannot be assumed to vanish for any $\Delta s \gg 0$.
As a consequence and in contrast to the Rouse model, the hydrodynamically coupled tension [Eq.~\eqref{eq:response_operator_zimm}],
\begin{equation}
    \hat{\mathcal{J}}\Delta R^2 = -\frac{D_0 \kappa}{L^2} \frac{\sqrt{6\pi}}{\Gamma(\nu) \cos\frac{\pi\nu}{2}} \, 
    \left(-L^2\partial_s^2\right)^{\frac{1+\nu}{2}} \Delta R^2 \, ,
\end{equation}
where I have used the semigroup property of the Riesz Laplacian, cannot self-tune to become marginal in steady state.
Instead, it must have a positive sign on all lengthscales, which is indeed the case for all $0 < \nu < 1$ as confirmed by substitution of the scaling ansatz,
\begin{equation}
    \hat{\mathcal{J}}\Delta R^2 = 2 D_0 \kappa \frac{\sqrt{6\pi} \Gamma(2\nu+1)}{\Gamma^2(\nu)} \tan\Bigl(\frac{\pi\nu}{2}\Bigr) \left|\frac{s-s'}{L}\right|^{\nu-1} .
\end{equation}
Incorporating hydrodynamically coupled self-interactions in the response operator, $\hat{\mathcal{J}} \to \hat{\mathcal{J}} + \hat{\mathcal{J}}_S$, proceeds analogously and leads to the typical scaling $\hat{\mathcal{J}}_S \Delta R^2 \sim \Delta s^{1-\nu + (n-1)(1-3\nu)}$ from the convolution of the $n$-body interactions with the Zimm kernel.

Figure~\ref{fig::scaling_active}(b) shows how these different physical mechanisms interact.
For monomers $s$ and $s'$ that are sufficiently close, $\Delta R(s,s') \ll \lambda$, the flow correlations saturate [Eq.~\eqref{eq:flow_correlations_asymptotic}].
For $|s-s'| \to 0$, the Neumann boundary condition dominates and forces the Rouse scaling, $\nu = 1/2$, which crosses over to an active shear-induced straightening of the polymer, $\nu = 1$.
For monomers that are sufficiently far apart, $\Delta R(s,s') \sim \lambda$, the flows begin to decorrelate according to Eq.~\eqref{eq:flow_correlations_asymptotic}, and the dynamics admits a new fixed point $\nu = 1/4$ balanced by polymer tension.
Such a hyper-compact scaling with fractal dimension $1/\nu = 4$ has been observed in chromatin~\cite{Article::Wang_2016} and prompted theoretical interest~\cite{Article::Shi_Thirumalai_2021, Chan_Rubinstein_2024, Article::Shi_Shin_Thirumalai_2025}.
The active Zimm theory recovers this anomalous scaling without invoking loop extrusion~\cite{Chan_Rubinstein_2024}.
Although the above scaling appears hyper-compact, it is an intermediate regime that appears directly after an intermediate swelling regime.
It cannot be valid for arbitrarily large distances $\Delta s$ in sequence space because states that are more compact than a globule are prohibited.
In agreement with this physical intuition, active flows decorrelate more steeply than thermal excitations for monomers that are far apart, which restores Rouse scaling even without invoking self-avoidance.

\section{Discussion}%
In summary, a main contribution of this work is a simple analytical framework based on a Gaussian approximation, which allows analyzing the conformational dynamics of active, hydrodynamically coupled, and self-interacting heteropolymers such as chromatin.
Compared to analogous work for passive polymers~\cite{Timoshenko_1995, Timoshenko_1995b, Timoshenko_1996, Timoshenko_1998, Jost_2014}, the derivation does not rely on the existence of a free energy describing the conformations of the polymer.
Therefore, the framework can be applied to active polymers that feature broken detailed balance and nonreciprocal interactions.
This was demonstrated for a scenario in which a chromatin locus produces RNA that attracts a different chromatin locus, but not vice versa~\cite{Article::Goh_2025}.

The unified framework geometrically interprets polymer folding as a non-linear and non-local diffusion of the pairwise squared separation (and thus contact probability) in sequence space (Fig.~\ref{fig::triangular_domain_diffusion}).
In locally homogeneous and approximately scale-free regions of the polymer, hydrodynamic interactions and self-interactions between monomers can be formally described by fractional calculus modeling anomalous diffusion.
This is analogous to how DNA coiling can facilitate protein target search by alternating between 1D diffusion along the polymer and 3D jumps mediated by looping~\cite{Berg_1981, Sokolov_1997, Lomholt_2005, Lomholt_2009, Benichou_2011}.
The fractional calculus formalism allows one to interrogate whether pairs of terms in the dynamical mean-field theory actually compete or not, beyond naively matching their exponents.

The theory was first validated by reproducing the known scaling behavior of self-avoiding chains and disordered globules.
The price of this simplicity is the neglect of non-Gaussian tails in the distributions of conformations.
This is internally consistent when frequent interactions lead to self-averaging, such as under melt conditions in the dense cell nucleus or for polymers collapsed to a globule with $\nu = 1/3$.
However, in dilute solution, it is well known that the prediction of Flory theory for the structural scaling exponent of self-avoiding chains, $\nu = 3/5$, which is close to the exact result $\nu = 0.588$, is based on a cancellation of errors.
It is therefore interesting that approximately Gaussian correlations are sufficient to reveal an upper bound, $\nu < 3/5$, because the conformation-averaged contact response [Eq.~\eqref{eq:interaction_pairwise_scalefree}] diverges logarithmically with chain length on further approach to the Flory fixed point.

The theory was then applied to active polymers under free-draining (Rouse) conditions and in a hydrodynamically coupled (Zimm) scenario.
Interestingly, analyzing the signs of the different terms in the dynamical mean-field theory, using the fractional calculus formalism, prohibited a fixed point where active hydrodynamic stirring (correlated flows) competes with tension on intermediate lengthscales for Rouse polymers, but allowed the same balance for Zimm chains.
In the latter scenario, for different loci $s$ and $s'$ that are further apart than the correlation length $\lambda$ of the flows, $\Delta R(s,s') > \lambda$, the theory predicts intermediate-scale hyper-compaction with fractal dimension $1/\nu = 4$ in actively driven Zimm chains.
Note that this fractal dimension cannot hold on arbitrarily large lengthscales and, indeed, the polymer returns to a Rouse-like structure for sufficiently large distances.

Active hydrodynamic stirring also causes an intermediate straightening regime on lengthscales where the distance between different loci $s$ and $s'$ is below the correlation length $\lambda$ of the flows, $\Delta R(s,s') \ll \lambda$.
This is analogous to the coil-to-stretch transition of polymers in turbulent flows~\cite{Review::Falkovich_2001}.
This swelling effectively renormalizes the Kuhn length and, conversely, it appears that nearby monomers, which receive correlated excitations, are more compactly organized than far-away monomers, which receive independent kicks.
This is in agreement with our previous work~\cite{Article::Goychuk2023}, where correlated hydrodynamic fluctuations effectively induced attraction between different genomic loci.
It will be interesting to explore how this couples to other mechanisms that can bring polymer loci together, such as correlated hydrodynamic fluctuations inducing capillary forces between biomolecular condensates~\cite{arxiv::Goychuk_2025} and chemical reactions mediating non-reciprocal interactions~\cite{Article::Goh_2025}.

At the level of contact maps, each nonequilibrium model studied in this work can be mimicked by a linear response theory in which the polymer folds due to additional Hookean springs~\cite{Article::Goychuk2023}.
However, they differ at the level of their response to perturbations and, moreover, have distinct experimentally testable signatures.
For example, the theory predicts that RNA synthesis at promoters, when combined with RNA gradient sensing at enhancers~\cite{Article::Goh_2025}, would lead to line-shaped features in contact maps [Figs.~\ref{fig::triangular_domain_diffusion}(c) and (d)].
Conversely, if digestion of RNA abolishes these features, it would be a strong indicator for a gradient-sensing mechanism.
The theory also predicts that, depending on the lengthscale, active shear flows lead to chromatin straightening and compaction [Fig.~\ref{fig::scaling_active}]; this could be tested experimentally by modulating cytoskeletal forces.

In future work, this framework can be further tested by studying the propagation of tension along chains that are inextensible~\cite{Article::Hallatschek2007}, incorporating hard confinement, and considering higher-order interactions.
For example, self-attraction overcoming tension and competing with bending could lead to Allen-Cahn or Cahn-Hilliard like dynamics~\cite{Bray_1994} and thus phase separation in sequence space, analogously to microphase separation in block copolymers~\cite{Leibler_1980}.
Spinodal decomposition during collapse has been studied for the equilibration of passive homopolymers and heteropolymers~\cite{Timoshenko_1995, Timoshenko_1996, Timoshenko_1998}, but the role of active processes is currently underexplored.
In this context, the present framework could directly enable studying the effect of heterogeneous active processes in the sequence and embedding space on, for example, multistability~\cite{Timoshenko_1998, Jost_2014}.
Moreover, future extensions could include non-Markovian effects due to active tension fluctuations~\cite{preprint::Ciarchi_2026}, viscoelastic memory in the surrounding solvent~\cite{Lampo_2016}, nonstationarity of concentration profiles, and self-propulsion of chemically active condensates~\cite{Article::Demarchi_Goychuk_2023, Article::Goychuk_Demarchi_2024, Qiang_2025} at specific genomic loci, which could lead to motility-induced phase separation~\cite{Cates_Tailleur_2015}.

More generally, mapping the problem of active polymer folding to non-local diffusion and reaction among pairwise contact probabilities opens a path towards the analytical study of the joint dynamics of chromatin-bound proteins and chromatin folding with the rich machinery of reaction-diffusion systems~\cite{Frey_Weyer_2026}.
In this context, the coupling between concentration fields and geometry~\cite{Wurthner_2023}, both along the chain and in the nucleoplasm, could be particularly promising.

\section*{Data Availability}
The simulations were implemented in Julia~\cite{software::julia} and the code is available on Zenodo~\cite{simulations}.

\begin{acknowledgments}
The author thanks Mehran Kardar, Arup K. Chakraborty, Deepti Kannan, Kirill Polovnikov, and Matteo Ciarchi for insightful discussions and feedback.
A.G. acknowledges support by the Ministry of Science and Culture of Lower Saxony through funds from the program zukunft.niedersachsen of the Volkswagen Foundation for the ‘CAIMed – Lower Saxony Center for Artificial Intelligence and Causal Methods in Medicine’ project.
This work was started at MIT, where A.G. was supported by an EMBO Postdoctoral Fellowship (ALTF 259-2022).
\end{acknowledgments}

\bibliography{polymers}

\appendix
\begin{widetext}
\section*{End Matter}

\renewcommand{\theequation}{S\arabic{equation}}
\setcounter{equation}{0}

%
First, write out Eq.~\eqref{eq:force_balance} component-wise with Latin indices indicating sequence space and Greek indices indicating real space,
\begin{equation}
\label{eq:force_balance_component}
    \partial_t r_i^\alpha(t) = M_{ij}^{\alpha\beta}(\{\vec{r}(t)\}) \Bigl[
    f_j^\beta(t) + \eta_j^\beta(t) \Bigr] + \xi^\alpha(\vec{r}_i(t),t) \, .
\end{equation}
As discussed in the main text, we consider timescales on which the random variables $\eta$ and $\xi$ have decorrelated, which implies that they should be seen as the long-timescale limit of correlated fluctuations.
Consequently, Eq.~\eqref{eq:force_balance_component} should be interpreted in the Stratonovich sense.
After multiplying Eq.~\eqref{eq:force_balance_component} with $r_n^\gamma(t')$ and averaging over the noise,
\begin{equation}
\label{eq:force_balance_component_corr_r}
    \partial_t \left\langle r_i^\alpha(t) r_n^\gamma(t') \right\rangle = 
    \left\langle \left[ M_{ij}^{\alpha\beta}(\{\vec{r}(t)\}) \, f_j^\beta(t) \right] r_n^\gamma(t') \right\rangle
    + \Bigl\langle M_{ij}^{\alpha\beta}(\{\vec{r}(t)\}) \, \eta_j^\beta(t) r_n^\gamma(t') \Bigr\rangle
    + \Bigl\langle \xi^\alpha(\vec{r}_i(t),t) \, r_n^\gamma(t') \Bigr\rangle \, .
\end{equation}

Apply the Novikov-Furutsu theorem~\cite{Article::Novikov_1965, Article::Furutsu_1963} to the first term on the right-hand side with $\vec{r}$ as random variable, and to the second term with $\vec\eta$ as random variable.
For the third term, also apply the Novikov-Furutsu theorem but take into account that the noise is evaluated at random locations.
Together,
\begin{multline}
    \partial_t \left\langle r_i^\alpha(t) r_n^\gamma(t') \right\rangle = 
    \int \! d\tau \, \left\langle 
    \frac{\delta}{\delta r_m^\sigma(\tau)}\left[ M_{ij}^{\alpha\beta}(\{\vec{r}(t)\}) \, f_j^\beta(t) \right] 
    \right\rangle
    \left\langle r_m^\sigma(\tau) r_n^\gamma(t') \right\rangle \\
    + \int \! d\tau \, \left\langle \frac{\delta}{\delta \eta_m^\sigma(\tau)} \left[ M_{ij}^{\alpha\beta}(\{\vec{r}(t)\}) \, r_n^\gamma(t') \right]\right\rangle 
    \left\langle\eta_m^\sigma(\tau) \eta_j^\beta(t) \right\rangle \\
    + \int \! d\tau \, \int \! d^3\vec{x} \, \left\langle \frac{\delta r_n^\gamma(t')}{\delta \xi^\beta(\vec{x},\tau)} \left\langle \xi^\beta(\vec{x},\tau) \xi^\alpha(\vec{r}_i(t),t) \right\rangle \right\rangle \, .
\end{multline}
Now, substitute the covariances of the two different noise terms defined in the main text, namely Eqs.~\eqref{eq:noise_covariance_monomers}~and~\eqref{eq:noise_covariance_spatial}.
Moreover, use spatial isotropy which in 3D indicates $\left\langle r_m^\sigma(\tau) r_n^\gamma(t') \right\rangle = \left\langle \vec{r}_m(\tau) \cdot \vec{r}_n(t') \right\rangle \frac{1}{3} \delta_{\sigma\gamma}$.
After contracting over the Greek indices,
\begin{multline}
    \partial_t \left\langle \vec{r}_i(t) \cdot \vec{r}_n(t') \right\rangle = 
    \int \! d\tau \, \frac{1}{3} \left\langle 
    \frac{\delta}{\delta r_m^\alpha(\tau)}\left[ M_{ij}^{\alpha\beta}(\{\vec{r}(t)\}) \, f_j^\beta(t) \right] 
    \right\rangle
    \left\langle \vec{r}_m(\tau) \cdot \vec{r}_n(t') \right\rangle \\
    + \frac{1}{3} \left\langle \frac{\delta}{\delta \eta_m^\beta(t)} \left[ M_{ij}^{\alpha\beta}(\{\vec{r}(t)\}) \, r_n^\alpha(t') \right]\right\rangle 
    C_{mj} 
    + \int \! d^3\vec{x} \, \frac{A}{3} \left\langle \frac{\delta r_n^\alpha(t')}{\delta \xi^\alpha(\vec{x},t)} 
    C_\xi(\vec{x}-\vec{r}_i(t))
    \right\rangle \, .
\end{multline}
Consider that the flow is incompressible, whereby the spatial divergence of the mobility tensor must vanish at any point in space, and that the deterministic forces are Markovian (that is, only depend on the current state and not on past times),
\begin{multline}
\label{eq:appendix:correlator_dynamics_intermediate}
    \partial_t \left\langle \vec{r}_i(t) \cdot \vec{r}_n(t') \right\rangle = 
    \hat{J}_{im}
    \left\langle \vec{r}_m(t) \cdot \vec{r}_n(t') \right\rangle \\
    + \frac{1}{3} \left\langle \frac{\delta}{\delta \eta_m^\beta(t)} \left[ M_{ij}^{\alpha\beta}(\{\vec{r}(t)\}) \, r_n^\alpha(t') \right]\right\rangle 
    C_{mj} 
    + \int \! d^3\vec{x} \, \frac{A}{3} \left\langle \frac{\delta r_n^\alpha(t')}{\delta \xi^\alpha(\vec{x},t)} 
    C_\xi(\vec{x}-\vec{r}_i(t))
    \right\rangle \, ,
\end{multline}
where the effective response matrix is given by
\begin{equation}
\label{eq:appendix:effective_response_matrix}
    \hat{J}_{im}(t) \coloneqq
    \frac{1}{3} \left\langle 
    M_{ij}^{\alpha\beta}(\{\vec{r}(t)\}) \, \frac{\delta f_j^\beta(t)}{\delta r_m^\alpha(t)}  
    \right\rangle \, .
\end{equation}
After applying the product and the chain rule to the second term on the right-hand side of Eq.~\eqref{eq:appendix:correlator_dynamics_intermediate},
\begin{multline}
    \partial_t \left\langle \vec{r}_i(t) \cdot \vec{r}_n(t') \right\rangle = 
    \hat{J}_{im}
    \left\langle \vec{r}_m(t) \cdot \vec{r}_n(t') \right\rangle 
    + \frac{1}{3} \left\langle \frac{\delta M_{ij}^{\alpha\beta}(\{\vec{r}(t)\})}{\delta r_p^\sigma(t)} \, \frac{\delta r_p^\sigma(t)}{\delta \eta_m^\beta(t)} \, r_n^\alpha(t') \right\rangle 
    C_{mj} 
    \\
    + \frac{1}{3} \left\langle M_{ij}^{\alpha\beta}(\{\vec{r}(t)\}) \, \frac{\delta r_n^\alpha(t')}{\delta \eta_m^\beta(t)} \right\rangle 
    C_{mj} 
    + \int \! d^3\vec{x} \, \frac{A}{3} \left\langle \frac{\delta r_n^\alpha(t')}{\delta \xi^\alpha(\vec{x},t)} 
    C_\xi(\vec{x}-\vec{r}_i(t))
    \right\rangle \, .
\end{multline}
The second term on the right-hand side can be further simplified by using $\delta r_p^\sigma(t) / \delta \eta_m^\beta(t) = M_{pm}^{\sigma\beta}$, which follows from Eq.~\eqref{eq:force_balance}, and applying the Novikov-Furutsu theorem~\cite{Article::Novikov_1965, Article::Furutsu_1963},
\begin{multline}
    \partial_t \left\langle \vec{r}_i(t) \cdot \vec{r}_n(t') \right\rangle = 
    \hat{J}_{im}
    \left\langle \vec{r}_m(t) \cdot \vec{r}_n(t') \right\rangle 
    + \frac{1}{3} \left\langle \frac{\delta}{\delta r_q^\gamma(t)} \left[ \frac{\delta M_{ij}^{\alpha\beta}(\{\vec{r}(t)\})}{\delta r_p^\sigma(t)} \, M_{pm}^{\sigma\beta}(\{\vec{r}(t)\}) \right]\right\rangle \left\langle r_q^\gamma(t) r_n^\alpha(t') \right\rangle 
    C_{mj} 
    \\
    + \frac{1}{3} \left\langle M_{ij}^{\alpha\beta}(\{\vec{r}(t)\}) \, \frac{\delta r_n^\alpha(t')}{\delta \eta_m^\beta(t)} \right\rangle 
    C_{mj} 
    + \int \! d^3\vec{x} \, \frac{A}{3} \left\langle \frac{\delta r_n^\alpha(t')}{\delta \xi^\alpha(\vec{x},t)} 
    C_\xi(\vec{x}-\vec{r}_i(t))
    \right\rangle \, .
\end{multline}
Continue manipulating this term by using spatial isotropy, whereby $\left\langle r_q^\gamma(t) r_n^\alpha(t') \right\rangle = \left\langle \vec{r}_q(t) \cdot \vec{r}_n(t') \right\rangle \frac{1}{3} \delta_{\gamma\alpha}$, and incompressibility of the fluid, whereby the spatial divergence of the mobility tensor must vanish at any point in space,
\begin{multline}
    \partial_t \left\langle \vec{r}_i(t) \cdot \vec{r}_n(t') \right\rangle = 
    \hat{J}_{im}
    \left\langle \vec{r}_m(t) \cdot \vec{r}_n(t') \right\rangle 
    + \frac{1}{9} \left\langle \frac{\delta M_{ij}^{\alpha\beta}(\{\vec{r}(t)\})}{\delta r_p^\sigma(t)} \, \frac{\delta M_{pm}^{\sigma\beta}(\{\vec{r}(t)\})}{\delta r_q^\alpha(t)} \right\rangle \left\langle \vec{r}_q(t) \cdot \vec{r}_n(t') \right\rangle 
    C_{mj} 
    \\
    + \frac{1}{3} \left\langle M_{ij}^{\alpha\beta}(\{\vec{r}(t)\}) \, \frac{\delta r_n^\alpha(t')}{\delta \eta_m^\beta(t)} \right\rangle 
    C_{mj} 
    + \int \! d^3\vec{x} \, \frac{A}{3} \left\langle \frac{\delta r_n^\alpha(t')}{\delta \xi^\alpha(\vec{x},t)} 
    C_\xi(\vec{x}-\vec{r}_i(t))
    \right\rangle \, .
\end{multline}
The theory can now be further simplified by eliminating the second term on the right-hand side in one of two equivalent ways.
Orientational averaging, $M_{ij}^{\alpha\beta} \to \tr(\mat{M}_{ij}) \delta_{\alpha\beta}/3$, which provides a decent approximation of hydrodynamic interactions~\cite{Tworek_Elcock_2023}, turns the gradients of the mobility tensor into divergences of the mobility tensor, which vanish due to fluid incompressibility.
The same simplification can be made without orientational averaging, by reasoning that $\delta M_{pm}^{\alpha\beta}/\delta r_q^\sigma(t) \sim \delta_{\alpha\sigma} \delta_{\beta\sigma}$, and also using fluid incompressibility.
Thus, one has
\begin{multline}
\label{eq:appendix:correlator_dynamics_intermediate_next}
    \partial_t \left\langle \vec{r}_i(t) \cdot \vec{r}_n(t') \right\rangle = 
    \hat{J}_{im}
    \left\langle \vec{r}_m(t) \cdot \vec{r}_n(t') \right\rangle  
    \\
    + \frac{1}{3} \left\langle M_{ij}^{\alpha\beta}(\{\vec{r}(t)\}) \, \frac{\delta r_n^\alpha(t')}{\delta \eta_m^\beta(t)} \right\rangle 
    C_{mj} 
    + \int \! d^3\vec{x} \, \frac{A}{3} \left\langle \frac{\delta r_n^\alpha(t')}{\delta \xi^\alpha(\vec{x},t)} 
    C_\xi(\vec{x}-\vec{r}_i(t))
    \right\rangle \, .
\end{multline}

Now consider the different temporal regimes of Eq.~\eqref{eq:appendix:correlator_dynamics_intermediate_next}.
Because of the intrinsic symmetry of the two-point correlator, $\left\langle \vec{r}_i(t) \cdot \vec{r}_n(t') \right\rangle$, focus on the domain $t \geq t'$ without loss of generality.
The second line of Eq.~\eqref{eq:appendix:correlator_dynamics_intermediate_next} indicates the response of the system at time $t'$ to perturbations applied at time $t$.
Causality whereby an event in the past cannot depend on the future implies
\begin{equation}
\label{eq:appendix:correlator_dynamics_positive_times}
    \partial_t \left\langle \vec{r}_i(t) \cdot \vec{r}_n(t') \right\rangle = 
    \hat{J}_{im}
    \left\langle \vec{r}_m(t) \cdot \vec{r}_n(t') \right\rangle
    \quad \text{for} \quad t > t' \, ,
\end{equation}
corresponding to Eq.~\eqref{eq:dynamic_mean_field_fluctuations} in the main text.
Thus, now only the initial condition at $t=t'$ needs to be determined.

When approaching $t = t'$ from $t' < t$, one has [Eq.~\eqref{eq:appendix:correlator_dynamics_positive_times}]
\begin{equation}
\label{eq:appendix:correlator_dynamics_positive_times_limit}
    \lim_{t'\xrightarrow{<} t} \partial_t \left\langle \vec{r}_i(t) \cdot \vec{r}_n(t') \right\rangle = 
    \hat{J}_{im}
    \left\langle \vec{r}_m(t) \cdot \vec{r}_n(t) \right\rangle
     \, .
\end{equation}
When approaching $t = t'$ from $t' \geq t$, one has [Eq.~\eqref{eq:appendix:correlator_dynamics_intermediate_next}]
\begin{multline}
\label{eq:appendix:correlator_dynamics_negative_times_limit}
    \lim_{t'\xrightarrow{>} t} \partial_t \left\langle \vec{r}_i(t) \cdot \vec{r}_n(t') \right\rangle = 
    \hat{J}_{im}
    \left\langle \vec{r}_m(t) \cdot \vec{r}_n(t) \right\rangle \\
    + \frac{1}{3} \left\langle M_{ij}^{\alpha\beta}(\{\vec{r}(t)\}) \frac{\delta r_n^\alpha(t)}{\delta \eta_m^\beta(t)} \right\rangle 
    C_{mj} 
    + \int \! d^3\vec{x} \, \frac{A}{3} \left\langle \frac{\delta r_n^\alpha(t)}{\delta \xi^\alpha(\vec{x},t)} 
    C_\xi(\vec{x}-\vec{r}_i(t))
    \right\rangle \, .
\end{multline}
Using the product rule, $\partial_t \langle \vec{r}_i(t)\cdot\vec{r}_n(t) \rangle = 
\lim_{t'\xrightarrow{>} t} \langle \partial_t\vec{r}_i(t)\cdot\vec{r}_n(t') \rangle + \lim_{t'\xrightarrow{<} t} \langle \vec{r}_i(t')\cdot \partial_t \vec{r}_n(t) \rangle$, and substituting Eqs.~\eqref{eq:appendix:correlator_dynamics_positive_times_limit}~and~\eqref{eq:appendix:correlator_dynamics_negative_times_limit} leads to
\begin{multline}
    \partial_t \langle \vec{r}_i(t)\cdot\vec{r}_n(t) \rangle =  
    \hat{J}_{im} \left\langle \vec{r}_m(t) \cdot \vec{r}_n(t) \right\rangle 
    + \left\langle \vec{r}_i(t) \cdot \vec{r}_m(t) \right\rangle \hat{J}^T_{mn} \\
    + \frac{1}{3} \left\langle M_{ij}^{\alpha\beta}(\{\vec{r}(t)\}) \frac{\delta r_n^\alpha(t)}{\delta \eta_m^\beta(t)} \right\rangle 
    C_{mj} 
    + \int \! d^3\vec{x} \, \frac{A}{3} \left\langle \frac{\delta r_n^\alpha(t)}{\delta \xi^\alpha(\vec{x},t)} 
    C_\xi(\vec{x}-\vec{r}_i(t))
    \right\rangle \, .
\end{multline}
The first term in the second line can be simplified using $\delta r_n^\alpha(t) / \delta \eta_m^\beta(t) = M_{nm}^{\alpha\beta}$, which follows from Eq.~\eqref{eq:force_balance}.
The second term in the second line can be simplified using $\delta r_n^\alpha(t) / \delta \xi^\gamma(\vec{x},t) = \delta(\vec{x} - \vec{r}_n(t)) \delta_{\alpha\gamma}$, which also follows from Eq.~\eqref{eq:force_balance}.
Together, this gives
\begin{multline}
    \partial_t \langle \vec{r}_i(t)\cdot\vec{r}_n(t) \rangle =  
    \hat{J}_{im} \left\langle \vec{r}_m(t) \cdot \vec{r}_n(t) \right\rangle 
    + \left\langle \vec{r}_i(t) \cdot \vec{r}_m(t) \right\rangle \hat{J}^T_{mn} \\
    + \frac{1}{3} \left\langle M_{ij}^{\alpha\beta}(\{\vec{r}(t)\}) M_{nm}^{\alpha\beta}(\{\vec{r}(t)\}) \right\rangle 
    C_{mj} 
    + A \left\langle   
    C_\xi(\vec{r}_n(t)-\vec{r}_i(t))
    \right\rangle \, .
\end{multline}
Finally, the covariance of the mobility tensor in the first term on the second line can be simplified either by using orientational averaging, or by reasoning that in an isotropic system the matrix $A^{\alpha\beta} = \langle M_{ij}^{\alpha\beta} M_{nm}^{\alpha\beta} \rangle$ can only have diagonal entries and that each diagonal entry must have the same value, $A^{\alpha\beta} = \frac{1}{9} \langle M_{ij}^{\gamma\gamma} M_{nm}^{\sigma\sigma} \rangle \delta_{\alpha\beta}$.
In summary, invoking isotropy essentially amounts to orientational averaging, and one has
\begin{equation}
    \partial_t X_{in}(t) =  
    \hat{J}_{im} X_{mn}(t) 
    + X_{im}(t) \hat{J}^T_{mn} 
    + \frac{1}{9} \left\langle \tr(\mat{M}_{ij}) \tr(\mat{M}_{nm})  \right\rangle 
    C_{mj} + A \left\langle   
    C_\xi(\vec{r}_n(t) - \vec{r}_i(t))
    \right\rangle \, ,
\end{equation}
where I have defined the same-time correlator $X_{in}(t) \coloneqq \langle\vec{r}_i(t) \cdot \vec{r}_n(t)\rangle$.
To deal with the spatial noise, consider the Taylor expansion
\begin{equation}
    \left\langle   
    C_\xi(\vec{r}_n - \vec{r}_i)
    \right\rangle = \sum_k \frac{1}{k!} \left\langle [(\vec{r}_n - \vec{r}_i) \cdot \vec\nabla]^k \right\rangle C_\xi(\vec{x} = 0) \, .
\end{equation}
After using Wick's theorem, and the independence of fluctuations along the $d=3$ orthogonal spatial directions, $\langle\Delta\vec{r} \otimes \Delta\vec{r}\rangle = \langle|\Delta \vec{r}|^2\rangle \, \mat{I} / 3$, one has
\begin{equation}
    \left\langle   
    C_\xi(\vec{r}_n - \vec{r}_i)
    \right\rangle = \sum_{k \, \text{even}} \frac{1}{k!} \frac{k!}{(k/2)! \, 2^{k/2}} \left[\frac{1}{3}\left\langle(\vec{r}_n - \vec{r}_i)^2 \right\rangle \vec\nabla^2\right]^{\frac{k}{2}} C_\xi(\vec{x} = 0) \, .
\end{equation}
This can be rewritten to
\begin{equation}
    \left\langle   
    C_\xi(\vec{r}_n - \vec{r}_i)
    \right\rangle = \sum_{k} \frac{1}{k!} \left[\frac{1}{6}\left\langle(\vec{r}_n - \vec{r}_i)^2 \right\rangle \vec\nabla^2\right]^{k} C_\xi(\vec{x} = 0) \, .
\end{equation}
Summation leads to 
\begin{equation}
\label{eq:appendix:exponential_laplacian}
    \left\langle   
    C_\xi(\vec{r}_n - \vec{r}_i)
    \right\rangle 
    = \exp\left[\frac{1}{6}
    \Delta R_{ni}^2 \vec\nabla^2\right] C_\xi(\vec{x} = 0) 
    = \bigl\langle C_\xi \bigr\rangle (\Delta R_{ni}) \, ,
\end{equation}
where I have assumed that the correlation function is isotropic, $C_\xi(\vec{x}) \equiv C_\xi(|\vec{x}|)$.

Finally, to simplify dealing with the hydrodynamic self-interactions coupling to the excitations, I apply pre-averaging of the Zimm kernel, 
\begin{equation}
    \mat{M}_{ij} \approx \mat{I} \, \hat{M}_{ij} \, , \quad \text{with} \quad \hat{M}_{ij} \coloneqq \frac{1}{3} \left\langle \tr(\mat{M}_{ij}) \right\rangle \, ,
\end{equation}
which corresponds to Eq.~\eqref{eq:zimm_kernel} in the main text.
This recovers Eq.~\eqref{eq:dynamic_mean_field_full} in the main text
\begin{equation}
    \partial_t X_{in}(t) =  
    \hat{J}_{im} X_{mn}(t) 
    + X_{im}(t) \hat{J}^T_{mn} 
    + \hat{M}_{ij} C_{jm} \hat{M}_{mn} + A 
    \left\langle C_\xi \right\rangle (\Delta R_{ni}) \, ,
\end{equation}
where I have used that $C_{ij}$ and $\hat{M}_{ij}$ are both symmetric.
Note that the pre-averaging approximation is only a choice to simplify the calculations, and not necessary to carry out the Gaussian averages.
Also applying the pre-averaging approximation of the Zimm kernel to the effective response matrix Eq.~\eqref{eq:appendix:effective_response_matrix} for consistency yields Eq.~\eqref{eq:effective_response_matrix} in the main text
\begin{equation}
    \hat{J}_{im}(t) \approx
    \hat{M}_{ij} \, \frac{1}{3} \left\langle 
    \frac{\delta}{\delta \vec{r}_m(t)}  \cdot \vec{f}_j(t) 
    \right\rangle \, .
\end{equation}

\end{widetext}

\end{document}